\documentclass[a4,11pt]{article}

\usepackage{graphicx}
\usepackage{amsfonts}
\usepackage{amssymb}
\usepackage{mathrsfs}
\usepackage{gensymb}
\usepackage{mathtools}
\usepackage{xcolor}
\usepackage{hyperref}
\usepackage[bottom]{footmisc}
\usepackage{tabularx,ragged2e,booktabs,caption}
\newcolumntype{C}[1]{>{\Centering}m{#1}}

\usepackage{booktabs,tabularx}
\usepackage{algorithm}
\usepackage{algorithmic}
\usepackage{tabularx}
\usepackage{arydshln}
\usepackage{microtype}
\usepackage{authblk}

\begin{document}

\newcommand{\fixme}[1]{ { \bf \color{red}FIX ME \color{black} #1 } }
\newcommand{\comment}[1]{ { \bf \color{blue}Comment: \color{black} #1 } }
\newcommand*{\affaddr}[1]{#1}
\newcommand*{\affmark}[1][*]{\textsuperscript{#1}}
\providecommand{\e}[1]{\ensuremath{\times 10^{#1}}}
\newcommand{\fig}{figures/}
\newcommand{\figOld}{figures_old/}
\newcommand{\figTmp}{figures_tmp/}

\title{Modelling of hydro-mechanical processes in heterogeneous fracture intersections using a fictitious domain method with variational transfer operators}
\date{}

\author[1]{Cyrill von Planta}
\author[2]{Daniel Vogler}
\author[2]{Xiaoqing Chen}
\author[1]{Maria G.C. Nestola}
\author[2,3]{Martin O. Saar}
\author[1]{Rolf Krause}

\affil[1]{\small Institute of Computational Science, Universit\`{a} della Svizzera Italiana, Lugano,  6900, Switzerland}

\affil[2]{Geothermal Energy and Geofluids Group, Institute of Geophysics, Department of Earth Sciences, ETH Zurich, 8092 Zurich, Switzerland}

\affil[3]{Department of Earth and Environmental Sciences, University of Minnesota,  MN 55455, Minneapolis,  USA}




\maketitle
\begin{abstract}
Fluid flow in rough fractures and the coupling with the mechanical behaviour of the fractures pose great difficulties for numerical modelling approaches, due to complex fracture surface topographies, the non-linearity of hydromechanical processes and their tightly coupled nature. To this end, we have adapted a fictitious domain method to enable the simulation of hydromechanical processes in fracture-intersections. The main characteristic of the method is the immersion of the fracture, modelled as a linear elastic solid, in the surrounding computational fluid domain, modelled with the incompressible Navier Stokes equations. The fluid and the solid problems are coupled with variational transfer operators. Variational transfer operators are also used to solve contact within the fracture using a dual mortar approach and to generate problem specific fluid meshes. With respect to our applications, the key features of the method are the usage of different finite element discretizations for the solid and the fluid problem and the automatically generated representation of the fluid-solid boundary.
We demonstrate that the presented methodology resolves small-scale roughness on the fracture surface, while capturing fluid flow field changes during mechanical loading. Starting with 2D/3D benchmark simulations of intersected fractures, we end with an intersected fracture composed of complex fracture surface topographies, which are in contact under increasing loads.  The contributions of this article are: (1) the application of the fictitious domain method to study flow in fractures with intersections, (2)  a mortar based contact solver for the solid problem, (3) generation of problem specific grids using the geometry information from the variational transfer operators. 
\\
\noindent
\textbf{Keywords:} Fluid flow, Fracture mechanics, Mortar, $L^2$-projection, Fictitious domain, Multibody contact problem, Hydro-mechanical coupling, Geothermal energy
\end{abstract}
\section{Introduction}
Fluid transport behaviour in fractures governs reservoir engineering applications such as enhanced geothermal systems, oil- and gas recovery or CO$_2$ sequestration \cite{rasmuson_1986,tester_2006,mcclure_2014b,amann_2018}, since fluid flow rates in fractures can be magnitudes higher than in the solid rock matrix. 
At the field and reservoir scale, the fracture geometry configuration of a fracture network determines preferential flow paths, as connected fractures serve as flow conduits \cite{zimmerman_1991,dreuzy_2012,ebigbo_2016,hobe_2018,SZV+19}. 
At the scale of individual fractures, complex fracture geometries yield heterogeneous flow fields which are influenced by tightly coupled physics such as hydro-mechanical (HM) processes \cite{bandis_1983,tsang_1984,rutqvist_2003,vogler_2016c,vogler_2017}. 
Although a number of studies focus on HM processes at the fracture network scale \cite{mcclure_2014a,mcclure_2014b,fu_2016}, and a lot of work has been performed on single fracture behavior, investigations on HM processes in rough fractures remain scarce. 

For the mechanical behaviour of a single fracture, complex fracture surface topographies result in non-linear fracture opening or closure behavior, if the fracture is subjected to normal or shear loading  \cite{bandis_1983,pyraknolte_2000,jiang_2006,matsuki_2008,tatone_2015,vogler_2018,kling_2018}. 
Specifically, increasing mechanical load normal to the fracture yields non-linear, convergent fracture closure behavior, where increasingly larger load increments have to be applied for a given closure increment \cite{bandis_1983,matsuki_2008,zangerl_2008,vogler_2016,vogler_2016b}. 
Changes of the mechanical loading distribution on the fracture alter contact area and the aperture distribution across the fracture plane, which will yield complex aperture fields  \cite{tsang_1987,nemoto_2009,vogler_2018}. 
As a result, fluid flow fields in these aperture fields become highly heterogeneous and change with the loading \cite{tsang_1984,brown_1987,zimmerman_1991,zimmerman_1996}.

Between the single fractures and fracture networks, fracture intersections offer insights into the transition of hydro-mechanical behaviour from single fractures to fracture networks. 
Previous work on fracture intersections focused mostly on fluid flow \cite{hull_1986,kosakowski_1999,zou_2017}, mixing phenomena \cite{berkowitz_1994,Stockman_1997,johnson_2001,johnson_2006,michalis_2009}, solute transport \cite{mourzenko_2002}, and multi-phase flow \cite{MANOORKAR_2016}. 
At the fracture network scale, the influence of local fracture intersections on global flow behaviour has also been studied specifically \cite{park_2001,park_2003}. 
At the scale of individual fracture intersections, studies have shown rough fracture apertures to yield complex fluid flow patterns across fracture intersections \cite{berkowitz_1994,johnson_2001,mourzenko_2002,johnson_2006}.

Due to the non-linearity of hydro-mechanical processes, complex fracture surfaces and corresponding aperture fields, studies on HM processes on rough fractures with small scale roughness are uncommon. 
Instead, many modelling approaches use simplifying assumptions, such as local parallel plate models, one directional coupling, non-coupled hydro-mechanics, or linear relations between normal stress and fracture closure ~\cite{barton_1985,watanabe_2008,nemoto_2009}. 
This work presents a novel approach, where we use ideas from fictitious domain (FD) methods to capture the HM behaviour of fractures without the simplification of fracture topography or physical processes. This paper expands on previous work \cite{PVN+18,PVC+19} by considering hydro-mechanical effects on fracture intersections, while incorporating contact mechanics and fluid flow on fracture surfaces with complex topographical patterns.

Fictitious domain methods, on which our approach is based, go back to \cite{Hym52} and \cite{Sau63}. The main characteristic of FD methods is to substitute the solution of a PDE on a complex domain by embedding or immersing it in a simpler domain, on which it is easier to solve the problem.  This was found to be particularly useful when encountering fluid structure interaction (FSI). There, fluid flow in complex geometries can be simulated by immersing a solid into the fluid and letting it interact with the simulation in the simple domain. This may be done in multiple ways, such as representing the solid as a force term in the fluid \cite{Pes02}, enforcing no slip boundaries at the fluid-solid boundary \cite{GPH+99,Yu05}, or enforcing them over the entire intersection of the solid and the fluid \cite{HGC+14,BG03}. Since the fluid and the solid problem can be solved separately, a plethora of combinations of solid and fluid formulations, mesh types, and couplings is possible. We refer the reader to our references and the articles cited therein for more information on the many variants, and to \cite{HWL12} or \cite{Ric17} for a broader overview of FSI methods.

To highlight one of the advantages of FD methods, we consider one major alternative class, so called boundary-fitted methods. Boundary fitted methods solve the fluid subproblem in a moving spatial domain. The Navier-Stokes equations are thereby formulated in an arbitrary Lagrangian Eulerian framework \cite{DHP+04,NGC+16,NFV+17} and the solid problem is typically analysed in a Lagrangian fashion. Although these approaches are known to produce accurate results at the interface between solid and fluid, the fluid grid may become severely distorted for scenarios that involve large displacements or contact. In these cases, even with remeshing, the numerical stability of the coupled problem and the accuracy of the solution can be affected.
Our type of FD methods on the other hand, avoids the need for the remeshing of the fluid mesh completely. Neither in the case of large solid displacements, nor in the case of contact, we we need to fear an ill-conditioned computational fluid domain, which is a great advantage for simulating large fractures, as it simplifies the setup of the simulations immensely.

FD methods are also flexible in as much as they enable different formulations of the solid and the fluid problem. This is demonstrated in this article, where we have extended the approach in \cite{NBZ+18_1,PVN+18,PVC+19} to include a mortar-based contact method \cite{PVZ+19} in the solid problem and an augmented Lagrangian method for the coupling in the fluid problem. We formulate the solid problem with linear elasticity and linearized contact conditions,  whereas for the fluid problem we use the incompressible Navier Stokes equations. Each problem is formulated on a different mesh. The solid and fluid problem are simulated separately and we solve the entire system in a staggered approach. Central to the approach is the coupling of the two problems with variational transfer operators. These operators allow us to transfer physical quantities between the non-matching  meshes of the fluid and solid problem in a stable way and are a development from \cite{KZ16}. Within this article we use variational transfer operators on two other occasions: To generate the fluid meshes, which need to be refined with respect to the fractures a priori to the simulation, and to assemble the mortar projection, which resolves the nonmatching nodes at the contact surfaces of the solid problem.

This article is structured as follows: In Section~\ref{sec:methods}, owing to their omnipresence in this methodology, we formulate the variational transfer operators and explain their discrete assembly. Then we introduce the governing equations, the discretizations and the coupling of the solid and the fluid problem. The methodology is evaluated in Section~\ref{sec:results}, first with 2D and 3D benchmark examples of intersecting fractures and finally with two intersecting fractures which are placed under an increasing load. 

\section{Method}
\label{sec:methods}

\begin{figure}[hbt!] 
\centering             
\includegraphics[width=0.7\columnwidth]{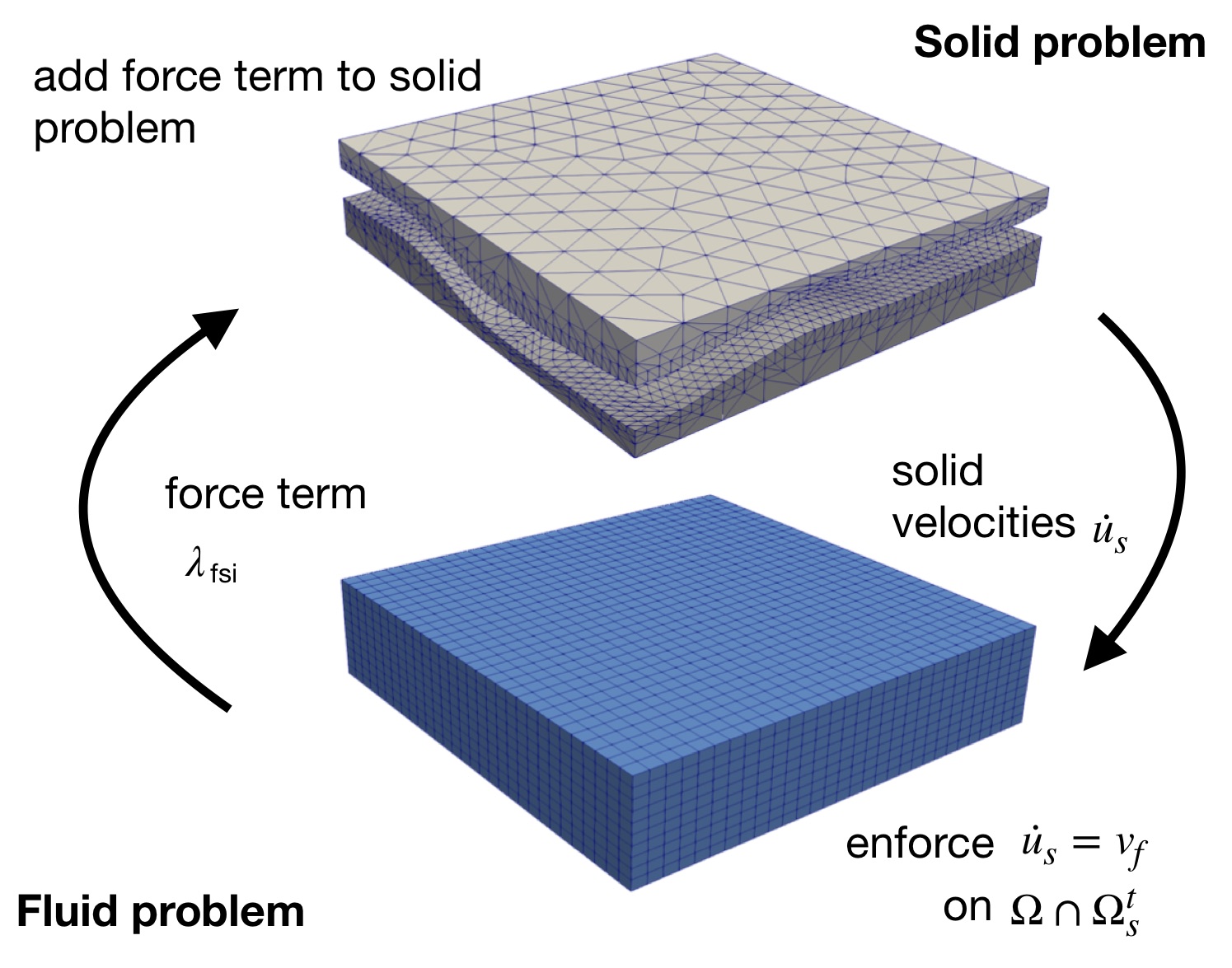} 
\caption{Sketch of the FD method. The fluid and solid interact by projecting solid velocities to the fluid and conversely by projecting force densities from the fluid to the solid.}
\label{fig:fsi_setup}
\end{figure}

The main idea of our fictitious domain method is, that the computational domain of the fluid also includes the solid. We use the incompressible Navier-Stokes equations to formulate the fluid problem and linear elasticity and linearized contact conditions for the solid problem. The fluid problem and the solid contact problem are solved separately with separate discretizations in a staggered approach. To couple the fluid-solid problem, we enforce the equality of fluid and solid velocities on the intersection of the solid and the computational domain of the fluid and transfer a fluid force term back to the solid problem (Fig.~\ref{fig:fsi_setup}). To this purpose we project the solid velocities from the solid to the fluid and conversely, project the force densities from the fluid to the solid. The projections themselves are so called variational transfer operators, or simply $L^2$-projections. The main advantage of variational transfer operators which we exploit here, is their ability to transfer quantities between different, non-matching meshes in a stable way. 

\begin{figure}[hbt!] 
\centering             
\includegraphics[width=0.5\columnwidth]{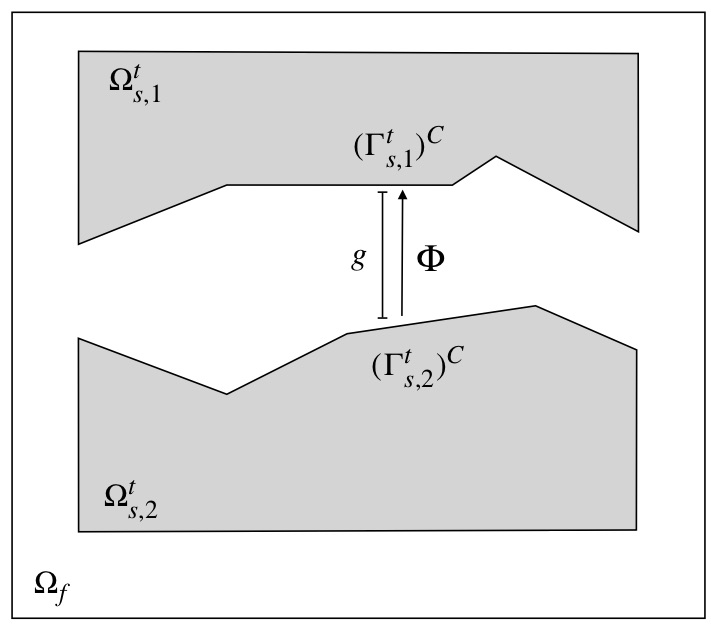}
\caption{Setup of the solid problem situated within the computational fluid domain $\Omega = \Omega_f \cup \Omega_s^t$, including the contact boundaries and gap function.}
\label{fig:fsi_contact}
\end{figure}
In the following sections we introduce the concept of variational transfer operators, followed by the strong and weak formulation of the FSI problem. We consider two non-intersecting solid bodies $\Omega^t_{s,1} $, $\Omega_{s,2}^t$, and write  $\Omega_f$ for the fluid. We use $\Omega_s^t$ for the union $\Omega^t_{s,1} \cup \Omega^t_{s,2}$, and $\Omega := \Omega^t_s \cup \Omega_f$ for the computational fluid domain, which is the union of the solid and the fluid. We use  $\Omega_{\text{fsi}}:=\Omega_{s}^t  \cap \Omega$ for the overlap of the computational fluid domain and the solid, which is used to formulate the coupling conditions in the weak formulation. Consequently we set  $\Gamma_{\text{fsi}} := \partial \Omega_{\text{fsi}}$. The boundary $\Gamma_f$ of the fluid is decomposed into a Dirichlet boundary $\Gamma_f^D$ and a Neumann boundary $\Gamma_f^N$, with $\Gamma_f = \Gamma_f^D \dot{\cup}  \Gamma_f^N$.  The solid bodies have the boundaries $\Gamma_{s,1}^t $ and $\Gamma_{s,2}^t$ and are decomposed into a Dirichlet, Neumann and contact boundary, such that $\Gamma_{s,\bullet}^t $=$   ( \Gamma_{s,\bullet}^t)^D \dot{\cup} ( \Gamma_{s,\bullet}^t)^N \dot{\cup} ( \Gamma_{s,\bullet}^t)^C$, $\bullet \in \{ 1, 2\}$. For ease of representation, we restrict ourselves to the formulation of a two-body contact problem here, whereby the indices $1$ and $2$ denote the mortar and non-mortar sides (Fig.~\ref{fig:fsi_contact}).
\subsection{Variational Transfer Operators} 
\label{sec:operators}
Let us assume that $V$ and $W$ are finite dimensional function spaces over the domain $\Omega$, which share the same $L^2$-scalar product $(f,g):= \int_{\Omega} f \, g \, d\omega$.  A variational transfer operator $\Pi$  maps elements from $V$ to $W$, while maintaining the equality of the elements in a weak, or variational sense for a yet to be defined multiplier space $M$:
\begin{eqnarray}
\Pi : V \longrightarrow W, \; v \mapsto w:= \Pi(v), \nonumber \\
\text{such that:}\; \int_{\Omega} (v - w) \mu \, d \omega =0, \, \forall \mu \in M. \label{eq_l2projection}
\end{eqnarray}
Since Eq.~\ref{eq_l2projection} represents an $L^2$-scalar product, mappings of this type are also called $L^2$-projections. To obtain a discrete representation of $\Pi$, we consider the bases $(\phi^{V}_i)_{i=1,...,n^{V}}$, $(\phi^{W}_j)_{j=1,...,n^{W}}$ and $(\phi^{M}_k)_{i=1,...,n^{M}}$ of $V,W$ and $M$ respectively, where $n^V$, $n^W$ and $n^M$ denote the dimension of each space and we demand that $n^W = n^M$. A discrete representation of $\Pi$ is obtained by replacing the elements in Eq.~\ref{eq_l2projection} with their basis representations $v = \sum_{i=1}^{n^V}v_i \phi_i^V$, $w = \sum_{j=1}^{n^W}w_j \phi_j^W$ and $\phi_k^M$:

\begin{eqnarray}
\sum_{i=1}^{n^V} v_i \int_{\Omega}  \phi^V_i  \phi^M_k \, d\omega = \nonumber \\ 
\sum_{j=1}^{n^W} w_j \int_{\Omega}  \phi^W_j \phi^M_k \, d\omega, \quad  k=1,...,n^W.
\label{eq_l2inter}
\end{eqnarray}
By defining the matrices $\mathbf{D}\in \mathbb{R}^{n^W\times n^W}$with entries
\begin{equation}
d_{jk} := \int_{\Omega}  \phi^W_j  \phi^M_k \, d \omega,
\end{equation}
$\mathbf{B} \in \mathbb{R}^{n^W\times n^V}$ with entries
\begin{equation}
b_{ik} := \int_{\Omega}  \phi^V_i  \phi^M_k \, d \omega,
\end{equation}
the coefficients of the basis representations of $v$ and $w$, the vectors $\mathbf{v}:=(v_i)_{i=1,...,n^V}$ and $\mathbf{w}:=(w_j)_{j=1,...,n^W}$, we get the discrete representation $\mathbf{T}$ of $\Pi$ from Eq.~\ref{eq_l2inter}, as $\mathbf{T}:=\mathbf{D}^{-1} \mathbf{B}$ and the following equalities:
\begin{equation}
\mathbf{B}\mathbf{v} = \mathbf{D} \mathbf{w} \Leftrightarrow \mathbf{D}^{-1} \mathbf{B} \mathbf{v} = \mathbf{w}  \Leftrightarrow \mathbf{T} \mathbf{v} = \mathbf{w}.
\end{equation}
In this form it may seem that a variational transfer operator is merely a change of basis for elements of finite element spaces with different bases on the same domain. We like to stress however, that the variational transfer operator provides the best approximation of elements $v \in V$ in $W$, in the sense of the $L^2$-scalar product, or, if we choose a multiplier space different from $W$, in a variational sense.

To avoid the computation of the possibly dense inverse of $\mathbf{D}$, we use here multiplier spaces $M$ that are spawned by biorthogonal basis functions \cite{Woh00}. This way, $\mathbf{D}$ becomes a diagonal matrix and the computation of its inverse straight forward (naturally, for other choices of $M$, one should ensure the regularity of $\mathbf{D}$). Note that the assembly of $\mathbf{T}$ essentially breaks down to computing the two mass matrices $\mathbf{D}$ and $\mathbf{B}$. On the implementation level, the main difficulty lies in the computation of the entries $b_{ik}$, where we need to determine the common support of the basis functions $\phi_i^V$ and $\phi_k^M$. These functions might be defined on different, nonmatching triangulations and in parallel computing environments, might even lie of different compute nodes. For more information on the parallel assembly algorithm we refer to  \cite{KZ16}.

$V$ and $W$ will be different finite element spaces in this article depending on the context of the application. For the coupling, they are finite element spaces defined on the fluid and the solid problem, and the resulting entries of the mass matrices are volume integrals.
For the contact problem, they are trace spaces defined on the boundaries of the solid bodies. In this case, the entries of the mass matrices will be surface integrals. We denote the exact choice of spaces when the discrete operators are effectively assembled or used.
%
\subsection{Governing equations}
In this section we introduce the fluid and the solid problem in its strong formulation. The fluid problem is given by the incompressible Navier-Stokes equations:
\begin{eqnarray}
\rho_f  \dot{v}  + \rho_f ( v \cdot \nabla) v  &\nonumber \\
 \qquad - \mu_f \nabla \cdot \sigma_f(p_f, v) =& f_f    \quad &\text{on} \; \Omega_f,  \label{eq_navier} \\
\nabla \cdot v =& 0   \quad & \text{on} \; \Omega_f, \label{eq_stokes} \\
v =& v_{D}  \quad & \text{on} \; \Gamma_f^D, \label{eq_bc_f_dc} \\ 
\sigma_f \cdot n_f =& h_f  \quad & \text{on} \; \Gamma_f^N. \label{eq_bc_f_nm}
\end{eqnarray}
Here, $\rho_f$ denotes the fluid density, $\mu_f$ the fluid's dynamic viscosity, $v$ the fluid velocity,  $\dot{v}$ the derivative of $v$ with respect to time, $p_f$ the fluid pressure, $f_f$ the forces acting on the fluid, $n_f$ the outer normal at the boundary, and $h_f$ the forces at the Neuman boundary. The fluid stress tensor $\sigma_f$ is given by $\sigma_f = -p_f \text{Id} + \frac{{\nabla v}^t + \nabla v} {2 }$.

The unconstrained solid problem is stated as:
\begin{eqnarray}
\rho_{s}  \ddot{u}  - \text{div}\sigma_s(u) &= f_s &\quad \mbox{on} \; \Omega^t_{s,1} \cup \Omega^t_{s,2}, \label{eq_eldynamics} \\ 
u &= u_{D} &\quad \mbox{on}\; (\Gamma_s^t)^D, \label{eq_solid_dc}\\ 
\sigma_s(u) \cdot n_s &= h_s &\quad \mbox{on} \; (\Gamma_s^t)^N.
\label{eq_solid_nm}
\end{eqnarray}
Here, $\rho_s$ denotes the solid density, $\dot{u}$ the solid velocity, $\ddot{u}$ the solid acceleration, $\sigma_s(u)$ the solid stress tensor according to Hooke's law, $f_s$ the body forces, $u$ the displacements, $n_s$ the outer normal at the boundary, and $h_s$ the forces at the Neumann boundary.

To formulate the contact conditions, we use the formulation from \cite{DK09} with slight adaptations. We assume a bijective mapping $\Phi$,
\begin{equation}
\Phi: (\Gamma_{s,2}^t)^C \rightarrow (\Gamma_{s,1}^t)^C
\label{eq_contact_phi}
\end{equation}
exists, which maps the points on the non-mortar side of the boundary to the possible contact point on the mortar side (Fig. \ref{fig:fsi_contact}). With $\Phi$ we define the vector field of normal directions $n^{\Phi}$:
\begin{eqnarray}
n^{\Phi} : (\Gamma_{s,2}^t)^C \longrightarrow \mathbb{S}^2, \;  n^{\Phi}(x)   := \left\{
                \begin{array}{lll}
                  \frac{\Phi(x)-x}{|\Phi{x}-x|}  &\quad & \text{if } \Phi(x) \neq x  \\
                  n_s(x) &\; & \text{otherwise}
                \end{array}
              \right .
\end{eqnarray}
The continuous gap function $g: \mathbb{R}^3 \longrightarrow \mathbb{R} \label{eq_g}, \,  x \mapsto | \Phi(x) -x |$ measures the width of the gap between the two bodies in the normal direction. $[u]  :=  (u_{2}  -  u_{1} \circ \Phi  )\cdot n^{\Phi}$ defines the point-wise jump of displacements $u_2$ on the slave and $u_1$ on the master side of the contact boundary. The jump has to be smaller than the gap $g$: $[u]  \leq  g$ and is to be interpreted pointwise. This condition is only meaningful in the linearized contact setting, where the bodies are close together and the outer normals $n^{\bullet}\, , \bullet \in \{1, 2\}$ are assumed to be parallel up to terms of higher order. The stresses and displacements with respect to the outer normal direction and the tangential direction $\top$ are  given by:
\begin{eqnarray}
\sigma_{n}^{\bullet} &= n_{s,i}^{\bullet} \cdot \sigma_s(u_{ \bullet})_{ij} \cdot n_{s,j}^{\bullet}, \quad  &u_{n,{\bullet}} = u_{\bullet} \cdot n_s^{\bullet},\\
\sigma_{\top}^{\bullet} &= \sigma_s(u_{\bullet})  n^{\bullet} - \sigma_n\cdot n^{\bullet}, \quad  &u_{\top}^{\bullet} = u_{\bullet} - u_{n,\bullet} \cdot n^{\bullet},
\end{eqnarray}
whereby we use the summation convention on repeated indices ranging from $1$ to $d$, with $d \in \{2,3\}$. We set $\sigma_n = \sigma_n^2$, $\sigma_{\top} = \sigma_{\top}^2$, and consequently the contact conditions read:
\begin{eqnarray}
\sigma_n &\leq&  0 \quad   \mbox{on} \quad (\Gamma_{s,2}^t)^C, \label{eq_neg_sigma_n} \\
\sigma_n (u_{1} \circ  \Phi)	&=&  \sigma_n (u_{2}) \quad \mbox{on} \quad (\Gamma_{s,2}^t)^C, \\
 \left[u \right]  &\leq& g \quad \mbox{on} \quad (\Gamma_{s,2}^t)^C,
 \label{eq_nonpen}\\
 \big( [u ] - g \big) \sigma_n(u_{2}) &=& 0 \quad \mbox{on} \quad (\Gamma_{s,2}^t)^C, \label{eq_compl}\\
\sigma_{\top} &=& 0 \quad \mbox{on} \quad (\Gamma_{s,2}^t)^C \label{eq_nofriction}.
\end{eqnarray}
Eq.~\ref{eq_nonpen} is the non-penetration condition, Eq.~\ref{eq_compl} are the complementary conditions and Eq.~\ref{eq_nofriction} states, that we set the tangential boundary stresses to zero, that is, we are solving for frictionless contact.

Finally, to couple the fluid and the solid problem, we enforce the equality of the fluid and solid velocity fields over the fluid-structure interface $\Gamma_{\text{fsi}}$ together with the traction forces:
\begin{eqnarray}
\label{eq_strong_coupling} 
\dot{u} &=& v, \\
\sigma_s n_s &=& \sigma_f n_f\; \quad \mbox{on} \quad \Gamma_{\text{fsi}}. \label{eq_strong_coupling_2} 
\end{eqnarray}
It is worth mentioning, that we could have also used a steady state formulation for the problem, considering the conditions we simulate.  However, to indicate the generality and extendability of the approach, we use time dependent formulations here.

\subsection{Weak Formulation} 

The weak formulation of the FSI problem consists of three parts. The first two are the weak formulations for the fluid and the solid problem, and the last one arises from the weak formulation of the coupling in Eq.~\ref{eq_strong_coupling}. Here we follow in large parts the example of \cite{HGC+14,NZK19} and more specifically \cite{KW12} for the solid part. Note however, that for the coupling of solid and fluid velocities, we rely on an augmented Lagrangian approach. We like to remark here, that an important change with respect to the strong formulation is, that the fluid problem will be defined on the computational domain $\Omega$, which includes the solid.  Thus, the Navier Stokes equations \eqref{eq_navier}-\eqref{eq_stokes} are also solved for the interior of the immersed solid. Furthermore, the coupling in Eq.~\ref{eq_strong_coupling} is enforced over $\Omega_{\text{fsi}}$ instead of $\Gamma_{\text{fsi}}$ only.

We begin by introducing the usual Sobolev spaces for the fluid velocities $v$, the fluid pressure $p$, and the solid displacement field $u$:
\begin{align}
\mathcal{W}^v &= \big \{\delta v \in [H^1(\Omega)]^d \, | \, \delta  v = 0 \, \mbox{on} \, \Gamma_f^D \big \}, \\
\mathcal{W}^p &=  [L^2(\Omega)]^d, \\
\mathcal{V}^{u} &= \big \{ \delta  u \in \prod_{i=1,2} \, [H^1(\Omega_{s,i}^t)]^d  \, | \, \delta  u=0\, \mbox{on} \, (\Gamma_s^t)^D \big \},
\end{align}
where $d \in \{2,3\}$. With test functions $\delta v \in \mathcal{W}^v$ and $\delta p \in \mathcal{W}^p$, we write the weak formulation of the fluid problem as:
\begin{align}
&\rho_f \int_{\Omega} \,  \big (\dot{v} +  (v \cdot \nabla \big ) \, v) \, \delta v \, dW \nonumber \\
&+  \int_{\Omega}    \sigma_f(p_f,v)  : \nabla ( \delta v) \, dW  +  
\int_{\Omega}  \delta p \nabla v \, dW \nonumber \\
&= \int_{\Gamma_f^N}  h_f  \cdot \delta v\, dA  +  \rho_f \int_{\Omega}   f_f \cdot \delta v  \, dW. \label{eq_fluid_weak}
\end{align}
Since $\Omega$ also comprises the solid $\Omega_s^t$, we obtain additional flow in the interior computational fluid domain. This interior flow field leads to additional fluid contributions, which in theory should be dealt with to maintain the force balance, as otherwise the condition Eq.~\eqref{eq_strong_coupling_2} might not be satisfied exactly. However, according to \cite{HGC+14}, Remark~2, these contributions can be disregarded in practice, because the contributions of the fluid stresses are much smaller than those of the solid.

For the solid problem we set $\Omega_s^t:= \Omega_{s,1}^t \cup \Omega_{s,2}^t$ and with test functions $\delta u \in \mathcal{V}^u$ we get the following weak formulation:

\begin{align}
\rho_s  \int_{\Omega_s^t}  \ddot{u} \cdot \delta u \, dV + \int_{\Omega_s^t} \sigma_s(u) : \nabla(\delta u)  \, dV \nonumber \\
- \int_{(\Gamma_{s}^t)^C} \sigma_n (u) \cdot [\delta u] \, dA    \nonumber \\
= \rho_s   \int_{\Omega_s^t}  f_s \cdot \delta u  \, dV - \int_{(\Gamma_{s}^t)^N}   h_s  \cdot \delta  u \, dA , \label{eq_solid_weak}
\end{align}
where the last term before the equality stands for the contact forces. These contact forces arise from the solution of the contact problem in the solid.  Since its static form, the weak formulation  of Eq.~\ref{eq_nonpen}, using a dual multiplier space $M_C$ \cite{Woh00}, reads as
\begin{equation}
\int_{(\Gamma_{s,2}^t)^C} ([u] - g) \cdot \lambda_C \, dA \leq 0. \quad \forall \lambda_C \in M_C,
\label{eq_solid_contact_weak}
\end{equation}
$\lambda_c$  will subsequently be interpreted as a force term and be used to formulate the linearized problem later in Section~\ref{sec_algorithm}.
To enforce the equality of solid and fluid velocities on the overlap $\Omega_{\text{fsi}}$ of the solid and fluid, we employ an augmented Lagrangian method by adding the following term (compare Eq.~\ref{eq_strong_coupling}):

\begin{align}
\int_{\Omega_{\text{fsi}}}  (v -\dot{u} ) \lambda_{\text{fsi}}\,+\,
\nonumber \\
\dfrac{\epsilon^{-1}}{2} \int_{\Omega_{\text{fsi}}} (v - \dot{u})^2 \,dV,\, \lambda_{\text{fsi}} \in M_{\text{fsi}}.
\label{eq_coupling_weak}
\end{align}
Here, $\epsilon$ is a penalty constant and $\lambda_{\text{fsi}} \in M_{\text{fsi}}$ a Lagrange multiplier. It is also possible to omit the penalty term in \eqref{eq_coupling_weak} and couple the problems using Lagrange multipliers only.

\subsection{Algorithm and Discretization}
\label{sec_algorithm}
\begin{figure}
\begin{center} 
\includegraphics[width=0.5\columnwidth]{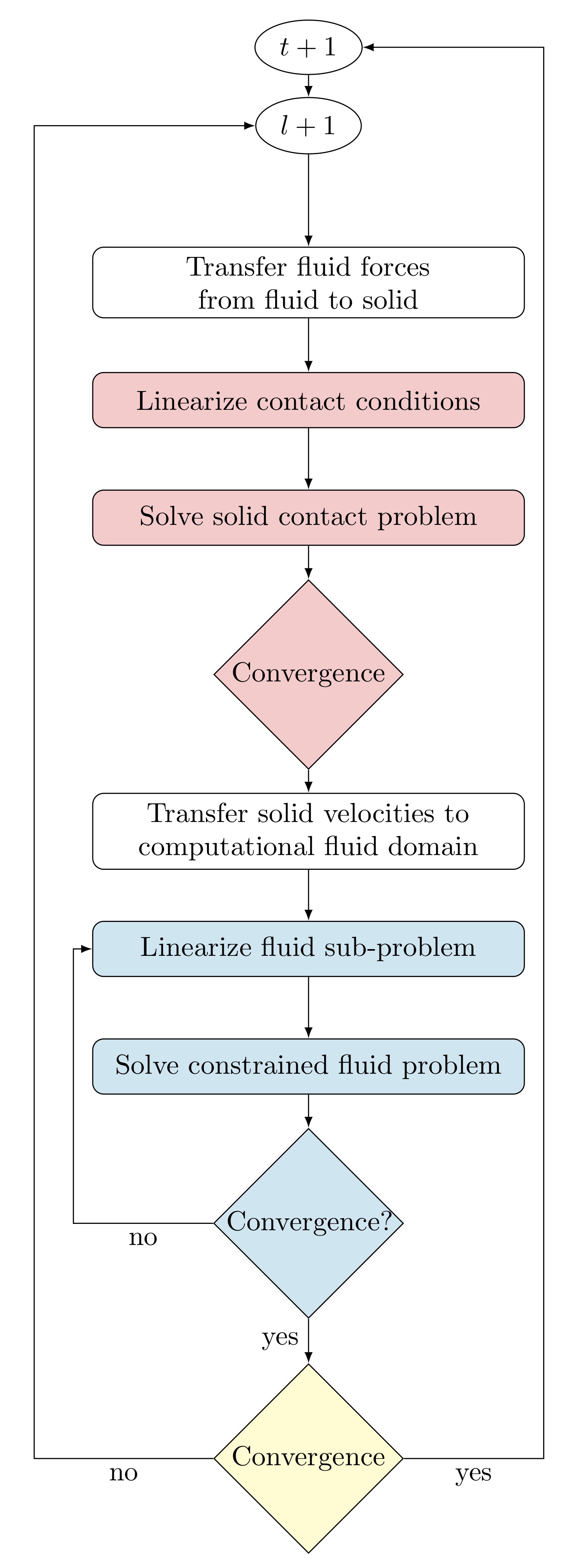}
\caption{Flowchart of FD method.}
\label{fig_flowchart}
\end{center}
\end{figure}

We solve the FSI problem in a staggered manner, whereby the solid and the fluid problem are solved sequentially until we reach convergence. We use finite elements to discretize both the fluid and the solid problem. 
Figure~\ref{fig_flowchart} shows a flowchart in which we depict one iteration $L$ for one time step $t$:
At the start of each iteration, we first transfer the fluid forces from the previous iteration to the solid, using variational transfer operators. In the next step we linearize the contact problem and solve the constrained minimisation problem, Eqs.~\ref{eq_solid_weak}-\ref{eq_solid_contact_weak}, with a semismooth Newton method.  
After this, we use variational transfer operators again, to transfer the solid velocities to the fluid problem (Eq.~\ref{eq_fluid_weak}). The fluid problem is solved with a nonlinear Newton method in several iterations, using SUPG and PSPG stabilization \cite{PLK+18}, whereby the weak constraints from Eq.~\ref{eq_coupling_weak} are enforced using an augmented Lagrangian approach. After the fluid problem has converged, we check for the overall convergence of the FSI problem and either move on to the next time step $t+1$ or the next iteration $l+1$. As measure for convergence we use the relative size of the step length of the solution to  a tolerance of $10^{-9}$.

To obtain discrete counterparts of $\mathcal{V}^{u}, \mathcal{V}^{\dot{u}},\mathcal{W}^{v}$ and $\mathcal{W}^{p}$, we discretize both the solid and the fluid problem with finite elements.
On the solid $\Omega_{s}^t$ we use unstructured meshes with tetrahedral elements and we denote the corresponding spaces of linear Lagrange elements for the solid displacements and velocities by $V^{u}$ and $V^{\dot{u}}$. 
The computational fluid domain $\Omega$ is meshed with hexahedral elements. We denote the corresponding space of bilinear Lagrange elements for the fluid velocities by $W^{v}$ and the space of bilinear Lagrange elements  for the fluid pressure with $W^{p}$.

For the linearized contact problem in the first part of the algorithm we define the operator $\mathcal{S}(u): \mathcal{V}^u \rightarrow (\mathcal{V}^u)^*$ as
\begin{eqnarray}
(\mathcal{S}(u), \delta u) := \rho_s \int_{\Omega_s}^t \ddot{u} \cdot \delta u \, dV + \int_{\Omega_s}^t \sigma_s(u) : \nabla(\delta u)  \, dV,
\label{eq_op_s}
\end{eqnarray}
where $(\mathcal{V}^u)^*$ denotes the dual space, and the right-hand side operator $\mathcal{R}_s(\delta u)$ as
\begin{eqnarray}
\mathcal{R}_s(\delta u) := \rho_s   \int_{\Omega^t_s}  f_s \cdot \delta u  \, dV - \int_{(\Gamma^t_s)^N}   h_s  \cdot \delta  u \, dA + \nonumber \\
\int_{\Omega_f \cap \Omega_s} \sigma_f  : \nabla ( \delta u) \, dV. \label{eq_rhs_s}
\end{eqnarray}
With $\mathbf{A}_s$ and $\mathbf{f}_s$ being the discrete variants of $\mathcal{S}$ and $\mathcal{R}_s$ respectively, the unconstrained solid problem can now be written as 
\begin{equation}
\mathbf{A}_s \mathbf{y}_s = \mathbf{f}_s.
\end{equation}
To formulate the linearized contact problem proper, we use a variational transfer operator $\mathbf{T}_C$, which relates the nodes from the master and slave side of the contact surfaces to each other.  In this context the operator is also called mortar operator. 
We denote the finite element spaces of the mortar and non-mortar body with $V^u_1$ and $V^u_2$, such that  $V^u = V^u_1 \times V^u_2$. The corresponding trace spaces on the contact boundaries $(\Gamma_{s,1}^t)^C$ and $(\Gamma_{s,2}^t)^C$ are denoted with  $\Gamma^u_1$ and $\Gamma^u_2$.
Using an approximation $\hat{\Phi}$ of the mapping $\Phi$ in Eq.~\ref{eq_contact_phi},
we form the mortar operator $\mathbf{T}_C: \Gamma_1^u \longrightarrow \hat{\Phi}(\Gamma_2^u)$ according to Section~\ref{sec:operators}. With  $\mathbf{u}_1$ and $\mathbf{u}_2$ being the coefficients of the vectors $u_1 \in \Gamma_1^u$ and $u_2 \in \Gamma_2^u$, and $\mathbf{g}$ being a discrete version of $g$, the  discrete contact condition from Eq.~\ref{eq_nonpen} becomes
\begin{equation}
\mathbf{u}_2 - \mathbf{T}_C \, \mathbf{u}_1  \leq \mathbf{g}.
\label{eq_contact_constr_discrete}
\end{equation}
The operator $\mathbf{T}_C$ is used to apply a change of basis on the assembled solid problem, which avoids the solution of a saddle point problem. Instead, one can apply a nonlinear solver such as monotone multigrid or semismooth Newton to solve the problem as a constrained minimization problem, yielding the previously mentioned Lagrange multiplier $\lambda_C$. For more details on the solution of the contact problem and its formulation, we refer the reader to \cite{WK03,DK09, Dic10,PVZ+19}.

The linearized fluid problem in the second part of the algorithm is formulated in a similar manner. We define the operator $ \mathcal{F}(v,p):= \mathcal{W}^v \times \mathcal{W}^p  \rightarrow (\mathcal{W}^v \times \mathcal{W}^p)^*$ as
\begin{eqnarray}
(\mathcal{F}(v,p), (\delta v, \delta p)) := \rho_f \int_{\Omega_f} \,  \big (\dot{v}_f 
+ (v \cdot \nabla  ) \, v \big) \, \delta v \, dW +   \nonumber \\
\int_{\Omega_f}    \sigma_f  : \nabla ( \delta v) \, dW + 
\int_{\Omega_f}  \delta v \nabla v \, dW \label{eq_F}
\end{eqnarray}
and the right hand side $\mathcal{R}_f(\delta v)$ operator as:
\begin{eqnarray}
\mathcal{R}_f(\delta v) := \int_{\Gamma_f^N}  h_f  \cdot \delta v\, dA  +  \rho_f \int_{\Omega_f}   f_{\text{fsi}} \cdot \delta v  \, dW. 
\label{eq_rhs_f}
\end{eqnarray}

As before we denote the discrete finite element variants of these operators with  $\mathbf{A}_f$ for $\mathcal{F}$ and $\mathbf{f}_f$ for $\mathcal{R}_f$. For the discrete representation of the coupling constraints in Eq.~\ref{eq_coupling_weak} we assemble the variational transfer operator $\mathbf{T}_{\text{fsi}}:= \mathbf{D}^{-1} \cdot \mathbf{B}:  V^{\dot{u}} \longrightarrow W^v$, again according to Sec.~\ref{sec:operators}. With $\mathbf{v}$ and $\mathbf{u}$ being the vectors of the basis representation of the fluid and solid velocity vectors, the discrete coupling condition reads:
\begin{equation}
\mathbf{v} = \mathbf{D}^{-1} \cdot \mathbf{B} \, \mathbf{u} \Leftrightarrow \mathbf{D} \, \mathbf{v}  =  \mathbf{B} \, \mathbf{u}.
\label{eq_coupling_discrete}
\end{equation}
To write the complete FSI problem in its linearized form, we introduce the penalty terms $\mathbf{P}_s$ and $\mathbf{P}_f$ originating from the second term in Eq.~\ref{eq_coupling_weak} of the augmented Lagrangian approach. We first define an operator $\mathbf{I}_{\Omega \cap \Omega_s^t}$ which determines the intersection of the solid domain and the computational fluid domain:
\begin{equation}
(\mathbf{I}_{\Omega \cap \Omega_s})_{kl} := \left\{
                \begin{array}{ll}
                1, \; \text{if}\, k=l, \text{and}\, \mathbf{T}_{\text{fsi}}(1) > t, \\
                0, \, \text{otherwise}.
                \end{array}
                \right.
\label{eq_indicator}
\end{equation}
By labelling the mass matrix obtained from the discretization on $W^v$with $\mathbf{M}_f$, we can write $\mathbf{P}_s$ as
\begin{equation}
\mathbf{P}_s := \epsilon^{-1} 
\left [
\begin{array}{c} 
\mathbf{I}_{\Omega \cap \Omega_s^t} \cdot \mathbf{M}_f \cdot \mathbf{T}_{\text{fsi}}  \\
\mathbf{0}
\end{array}
\right],
\end{equation}
and $\mathbf{P}_f$ as
\begin{equation}
\mathbf{P}_f := \epsilon^{-1} \left[
\begin{array}{cc}
\mathbf{I}_{\Omega \cap \Omega_s^t} \cdot \mathbf{M}_f & \mathbf{0}\\
\mathbf{0} & \mathbf{0}
\end{array}
\right],
\end{equation}
whereby the matrices $\mathbf{0}$ contain all the entries associated with the degrees of freedom from the fluid pressure.
The entire discretized solid-fluid problem can then be represented in the following block form:
\begin{equation}
\begin{array}{c} 
\text{(S)} \\
\text{(F)} \\
\text{(C)}
\end{array}
\left [
\begin{array}{ccc} 
\mathbf{A}_s + \mathbf{P}_s 	& - \mathbf{P}_f&\mathbf{B}^t \\ \hdashline
\mathbf{P}_s & \mathbf{A}_f - \mathbf{P}_f & \mathbf{D}^t \\
\mathbf{B}& \mathbf{D}&0
\end{array}
\right] 
\left[
\begin{array}{c}
\mathbf{y}_s^l \\ \hdashline
\mathbf{y}_f ^l\\
\mathbf{y}_{\lambda}^l
\end{array}
\right]
=
\left[
\begin{array}{c}
\mathbf{f}_s  + \mathbf{\lambda}_C \\ \hdashline
\mathbf{f}_f \\
0
\end{array}
\right].
\label{eq_discrete_system}
\end{equation}
With regard to the flowchart from the beginning of this section, the solid contact problem, which is solved at the start of the algorithm, is $(\mathbf{A}_s  - \mathbf{P}_s) \cdot \mathbf{y_s}^l = \mathbf{f}_s^{l-1} + \mathbf{\lambda}_C + \mathbf{P}_f \mathbf{y}_f^{l-1} - \mathbf{B}^t \mathbf{y}_{\lambda}^{l-1}$. Here, $\lambda_{C}$ denotes the multiplier which arises from the solution of the constrained minimization problem and $\mathbf{B}^t\, \mathbf{y}_{\lambda}^{l-1}$ has the interpretation of the force term mentioned in Fig.~\ref{fig:fsi_setup}.
After updating the solution vector $\mathbf{y}_s^l$ of the solid, we solve the constrained fluid problem, which corresponds to the rows F and C in Eq.~\ref{eq_discrete_system}, in several iterations. Thus, the algorithm corresponds to a block Gauss-Seidel scheme.  Note, that we did not introduce here the indices related to the time discretization $t$. For the fluid problem, we use a backward differentiation formula, BDF2, and for the solid problem the effective velocities turned out to be negligible in our numerical experiments, hence we can solve it as quasi-static problem.
\subsection{Implementation}
For the assembly of the discrete operators $\mathbf{A}_s$, $\mathbf{A}_f$, $\mathbf{P}_f$, $\mathbf{P}_s$, $\mathbf{f}_s$, and  $\mathbf{f}_f$ we use the MOOSE framework \cite{GNH+09}. The variational transfer operators $\mathbf{T}_{\bullet},\, \bullet \in \{ C, \text{fsi}\}$, as well as $\hat{\Phi}$ and $\mathbf{g}$ are implemented as user components within MOOSE, employing MOONoLith \cite{KZ16}, libMesh \cite{KPH+06} and Utopia \cite{ZKN+16}. In the same manner we implemented a specific, mortar-based contact solver in PETSC \cite{AdaAl2015} for the solid problem \cite{PVZ+19}.
\section{Numerical experiments}
\label{sec:results}
We apply the methodology to coupled fluid flow problems with mechanical contact to a series of problem settings with increasing complexity.  We use the same material properties for all simulations:  The linear elastic material is representative of hard granodiorite rock, with a Poisson ratio of 0.33, Young's modulus of 1\e{10}~Pa, solid rock density of 2.75~$g/cm^3$,  kinematic water viscosity of 1~$mm^2/s$, and a water density of 1~$g/cm^3$, for water at about 20$^\circ$C.  As these conditions reflect non-turbulent flow and the velocities in the solid are very low, the results represent quasi-static solutions. 
All simulations were ran until the solution of the fluid problem reached steady state and conducted on the computing cluster of the Institute of Computational Science in Lugano, Switzerland. 
Individual simulations used up to 10 compute nodes (2~x Intel Xeon E5-2650~v3 @~2.30GHz) with 16 CPU's each.
\subsection{Benchmark problems}
\label{sec:benchmarks}
We designed 2D and 3D benchmark simulations to assess the method's capability of simulating flow in geometries which contain multiple, intersecting flow channels. We apply a fluid pressure gradient of 1.0~Pa from the channels on the left to the single channel on the right, leading to flow through the intersection of the channels located in the center of the problem geometries. For the simulations conducted with our FD method, we apply zero-Dirichlet boundary conditions on the entire solid boundary, thereby creating a rigid object. This allows us to compare the simulation results with the solutions of an alternative Navier Stokes setup, where we employ only the equations for incompressible Navier-Stokes flow and a fluid geometry whose boundaries represent the same rigid solid. Aside from that, the experiments conducted with this alternative Navier-Stokes approach have the same setup as the simulations conducted with our FD method.

One of the drawbacks of our method is, that in the computational fluid domain, the nodes  in the intersection $\Omega_{\text{fsi}}$ are of little interest. Yet they occupy memory and computational resources. As a heuristic countermeasure we reduced their number as much as possible, by  generating all fluid meshes with selective refinement in regions of interest, that is around the fractures area. This is achieved by using the variational transfer operator from Sec.~\ref{sec:operators}. Starting with a relatively coarse mesh for the computational fluid domain, we first transfer an indicator variable from the solid to the fluid similar to Eq.~\ref{eq_indicator}. On the mesh of the computational fluid domain, the projected indicator is then either very small or zero, whenever there is little or no overlap with the solid. 
We use this information to refine the mesh on $\Omega$, wherever the value of the indicator variable on the nodes is smaller than the threshold or zero, indicating the boundary or interior of the fracture. 
With this we obtain a higher resolution within and around the fracture area, while using fewer nodes and thus less memory and computational effort in the less interesting areas.
\subsubsection{2D Benchmark problem}
\begin{figure*}
\begin{center}  
\includegraphics[width=0.95\textwidth]{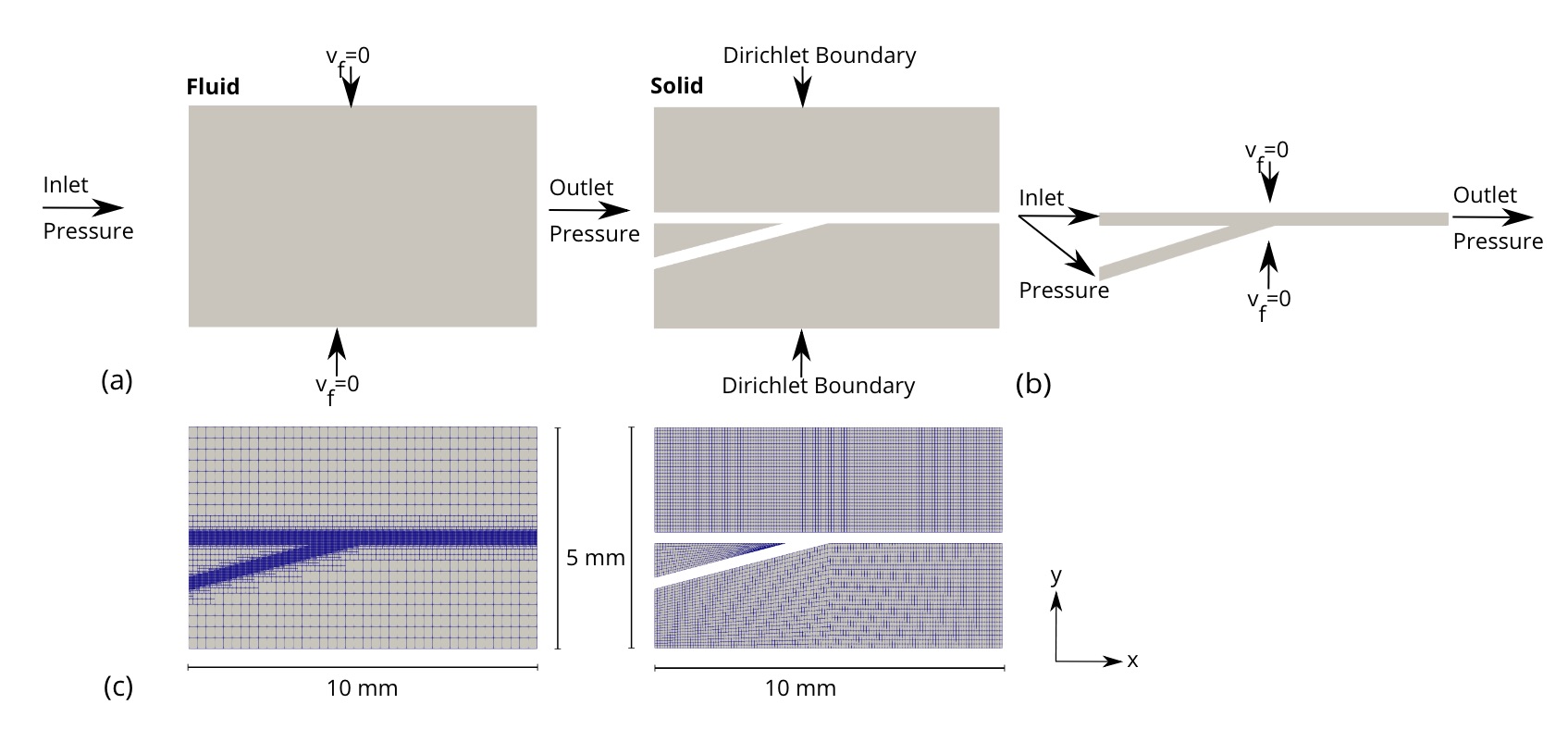}
\caption{Boundary conditions of the 2D benchmark case: a) Fluid and solid of the FD method; b) Alternative Navier-Stokes simulation; and (c) Fluid (left) and solid (right) meshes for 2D intersected-channel flow for the FD method.}
\label{fig:bnmrk_2d_bc_mesh}
\end{center}
\end{figure*}

The first benchmark problem simulates flow in a two-dimensional fracture with one intersection. It has two shorter fractures on the left and one longer fracture on the right side. 
The immersed solid that defines the channels, consists of an upper plate, a lower plate, and a wedge-shaped plate at the left side (Fig.~\ref{fig:bnmrk_2d_bc_mesh}a). 
We set the fluid velocity to zero along the lateral boundaries and apply a pressure gradient to the fluid mesh. This results in fluid flow through the two shorter channels, which merges into the single channel on the right. The solid  is immersed in a fluid with dimension 10~mm$\times$5~mm (Fig.~\ref{fig:bnmrk_2d_bc_mesh}c). 
The fluid mesh has 83'211 nodes with a mesh width $h_f$ of 0.0125~mm in the refined area within the fractures (Fig.~\ref{fig:bnmrk_2d_bc_mesh}c). The solid mesh consists of 8'419 nodes with a mesh width $h_s$ of about 0.1~mm.  The alternative Navier-Stokes setup has equivalent boundary conditions (see Fig.~\ref{fig:bnmrk_2d_bc_mesh}b). 
 
\begin{figure*}
\begin{center}      
\includegraphics[width=0.95\textwidth]{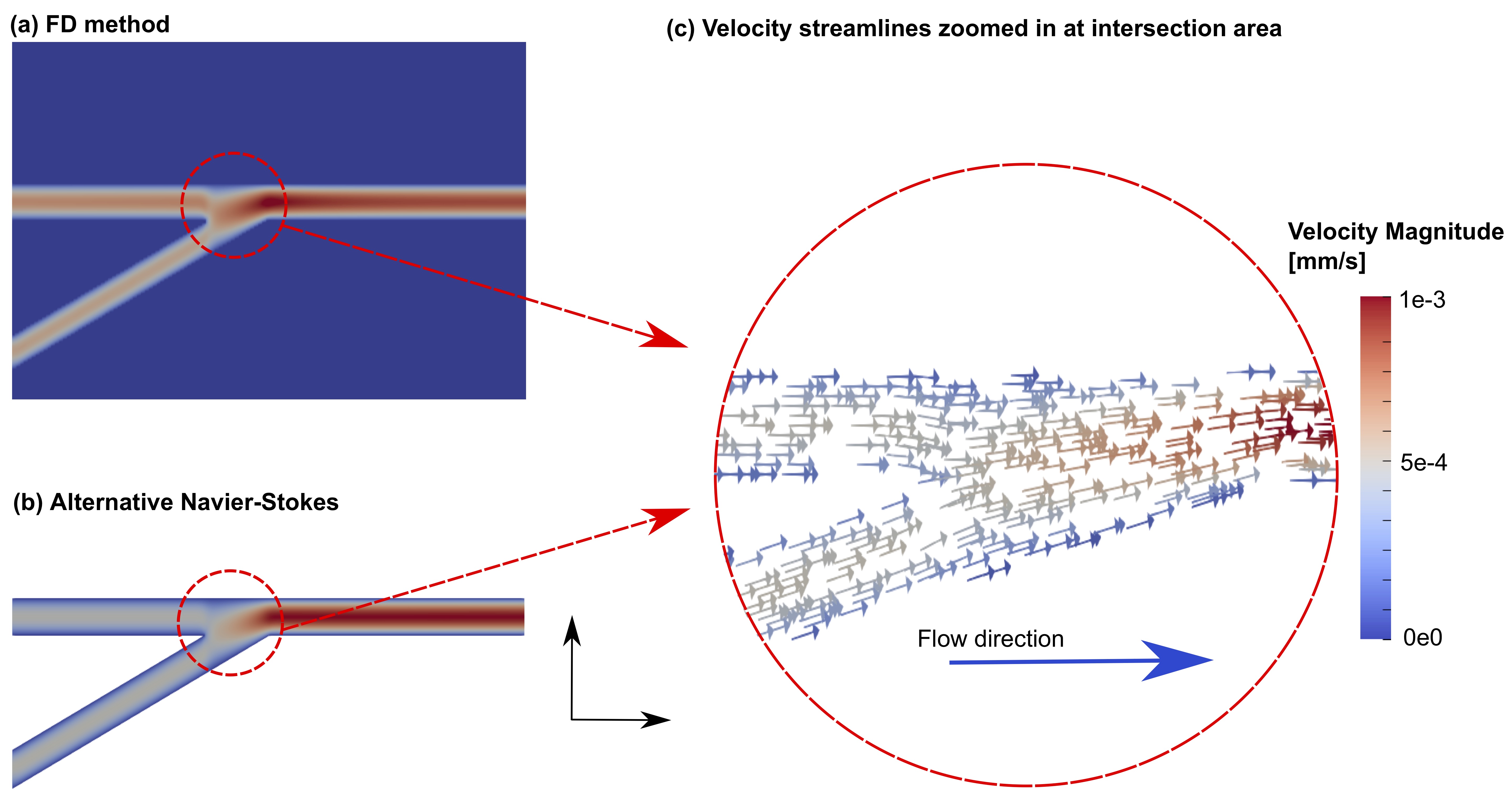}
\caption{Velocity distributions of: a) FD method; b) Alternative Navier-Stokes setup; and c) The velocity streamlines at the intersection area. }
\label{fig:bnmrk_2d_vel}
\end{center}
\end{figure*}

The results for the benchmark are shown in Fig.~\ref{fig:bnmrk_2d_vel}. They show good agreement between our FD method (Fig.~\ref{fig:bnmrk_2d_vel}a) and the alternative Navier-Stokes setup (Fig.~\ref{fig:bnmrk_2d_vel}b). 
Note that, as the fluid velocity is defined in the entire fluid domain, our FD method represents the solid as a large region of close-to-zero flow velocity (Fig.~\ref{fig:bnmrk_2d_vel}a). 
Fig.~\ref{fig:bnmrk_2d_vel}c shows the fluid velocity vectors in an enlarged region of the fracture intersection, where we can observe how flow from the two channels merges and the fluid velocity increases as a consequence. 

\subsubsection{3D Benchmark problem}
\begin{figure*}
\begin{center}  
\includegraphics[width=0.95\textwidth]{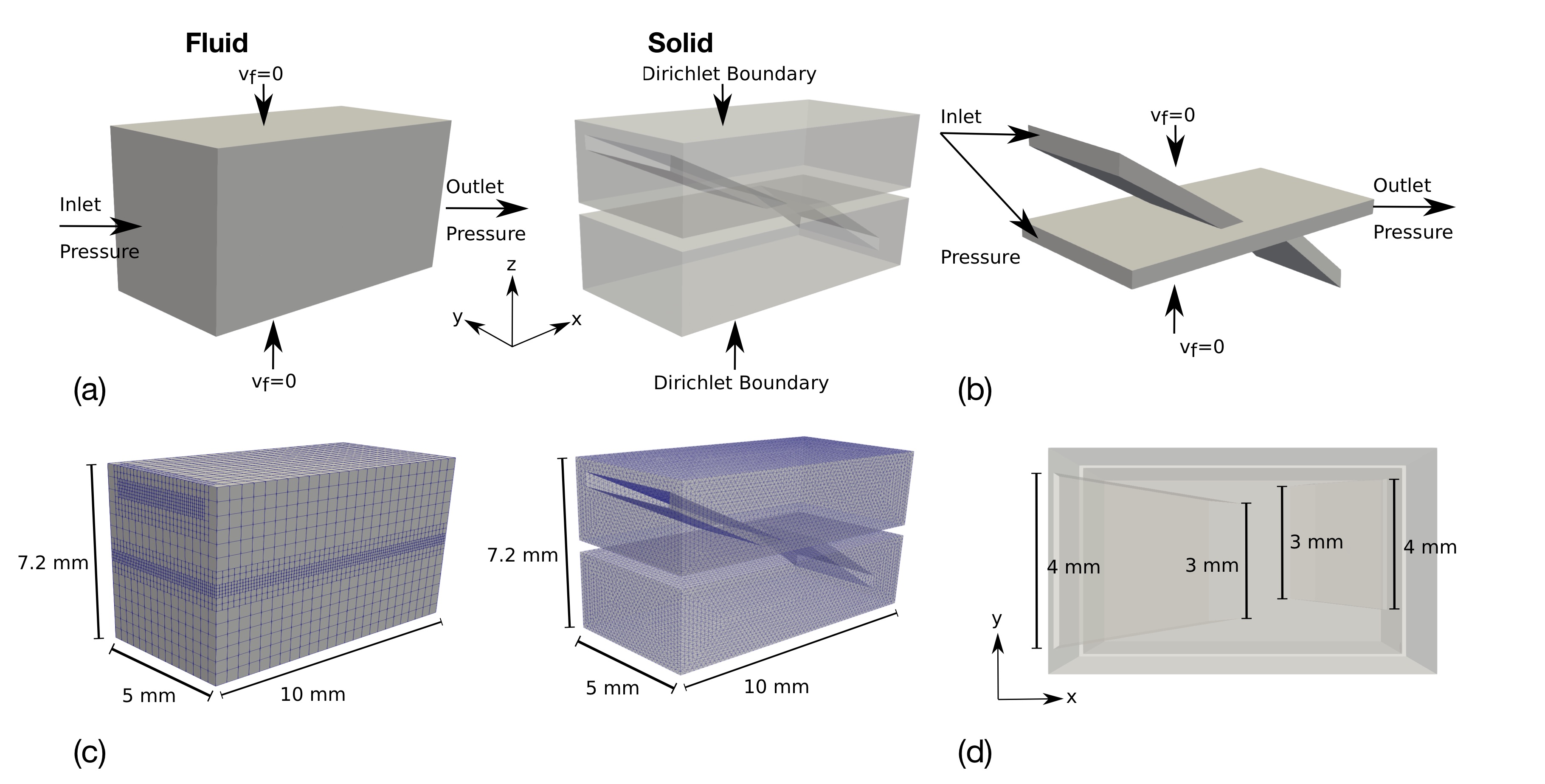}
\caption{Setup of the 3D benchmark simulation: a) Fluid and solid geometry used in the FD method; b) Fluid geometry and setup of alternative Navier-Stokes simulation; c) Fluid (left) and solid (right) meshes for FD method; and d) Top view on the intersecting flow channels.}
\label{fig:bnmrk_3d}
\end{center}
\end{figure*}

In this section we extend the previous benchmark to three dimensions (Fig.~\ref{fig:bnmrk_3d}). 
The angles between the main channel and the upper and lower flow channels are 30 and 40~degrees, respectively.
The other measures and boundary conditions are depicted in Fig.~\ref{fig:bnmrk_3d}. 
The fluid mesh has 534'455 nodes, with a  mesh width $h_f$ of 0.05~mm in the area where flow is expected. The outer boundaries of the solid, which define the channels, have the same length, width, and height as the fluid domain. 
The unstructured solid mesh has 120'159 nodes and a mesh width $h_s$ of roughly 0.15~mm (see Fig.~\ref{fig:bnmrk_3d}c). 

\begin{figure*}
\begin{center}  
\includegraphics[width=0.95\textwidth]{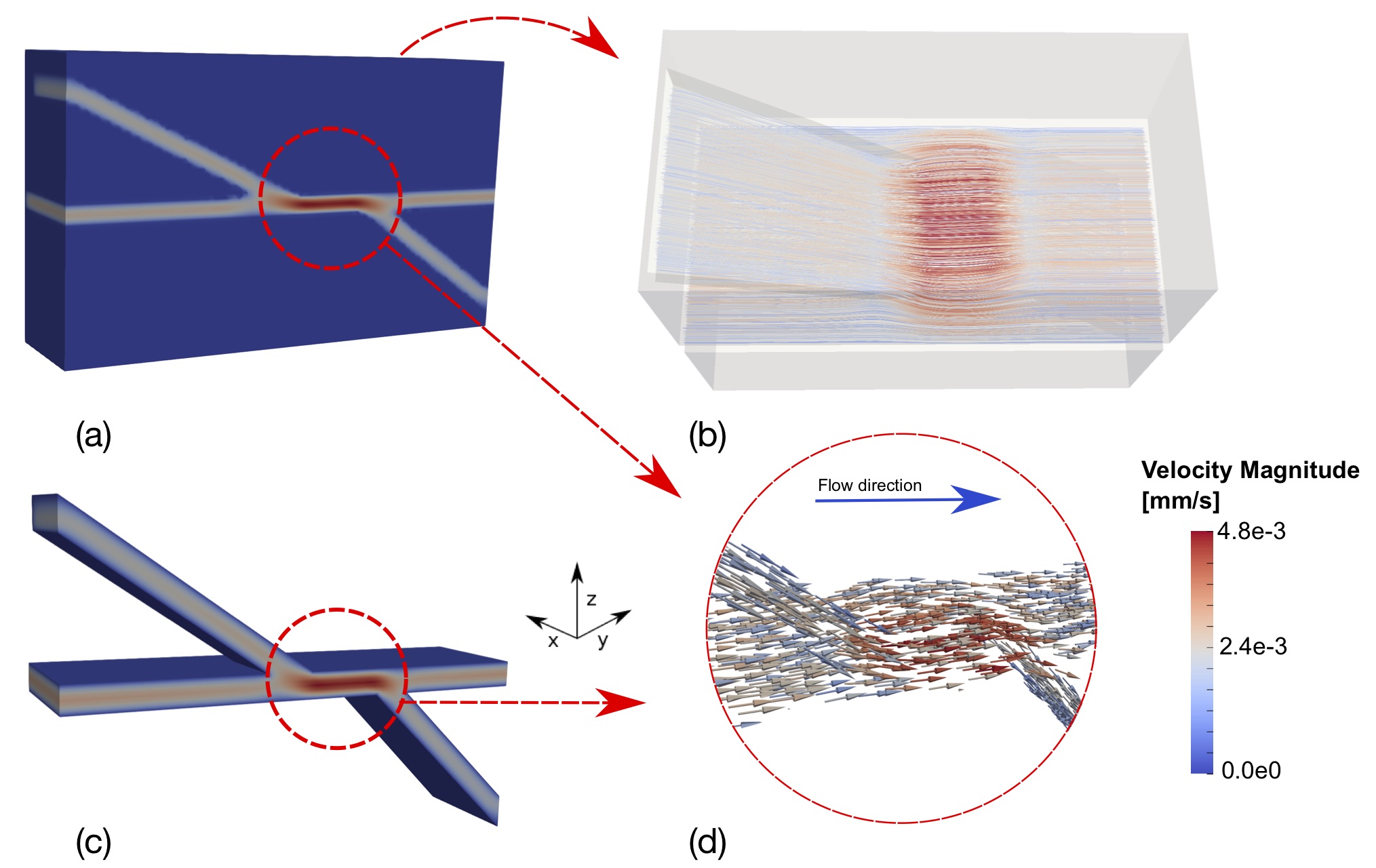}
\caption{a) Velocity magnitude of fluid from the FD method; b) Velocity magnitude of fluid for the alternative Navier-Stokes simulation; c) Velocity streamlines of fluid flow; and d) Velocity vector field at the intersection regions. }
\label{fig:bnmrk_3d_vel}
\end{center}
\end{figure*}

\begin{figure*}
\begin{center}  
\includegraphics[width=0.7\columnwidth]{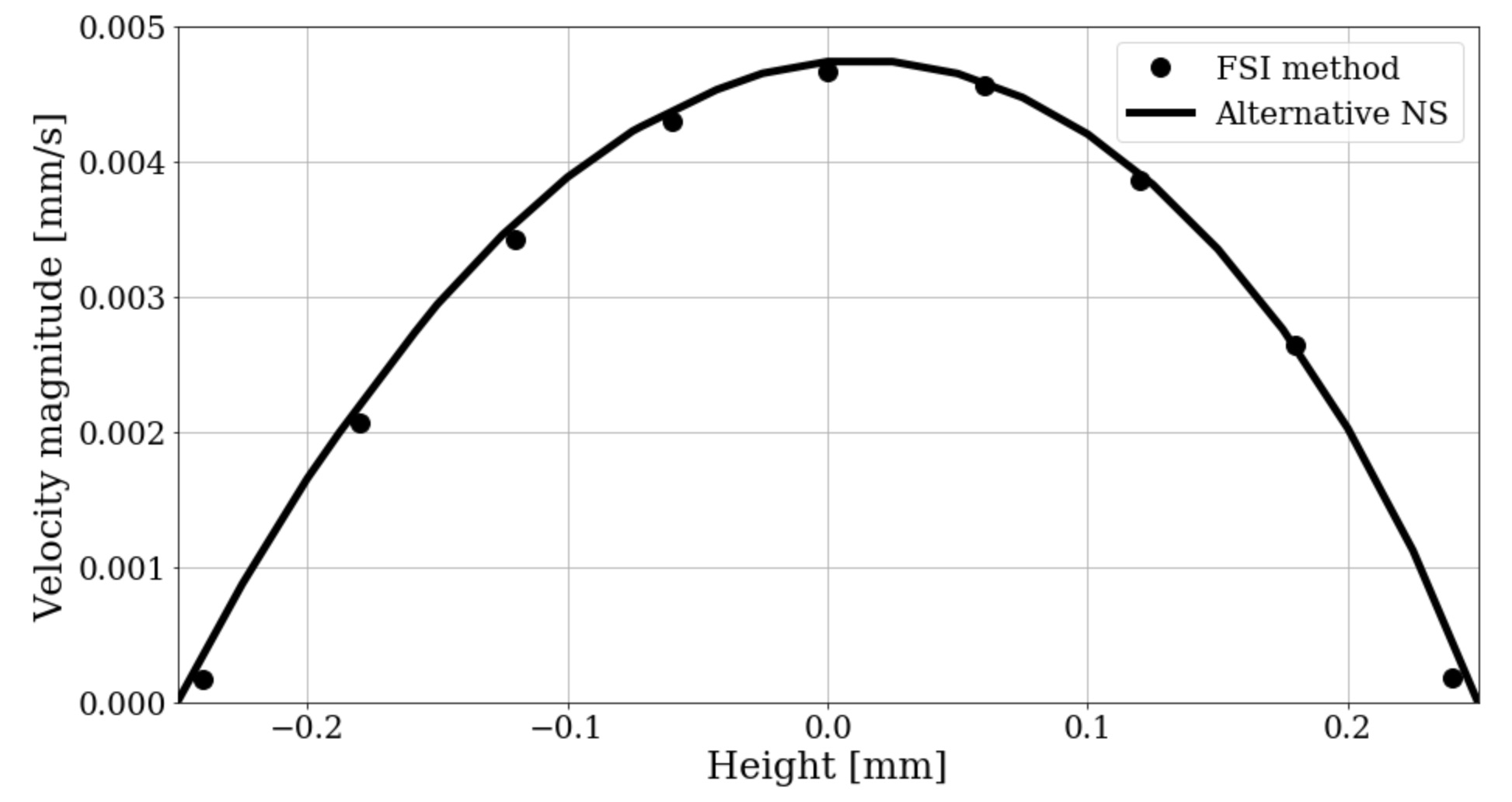}
\caption{Velocity profiles at x=5~mm and y=2.5~mm of the 3D benchmark simulation,comparing solution of FD method with the alternative Navier-Stokes approach.}
\label{fig:bnmrk_3d_cmp}
\end{center}
\end{figure*}

We use this problem first to quantitatively asses the validity of the FSI method. In Fig.~\ref{fig:bnmrk_3d_cmp} we compare the velocity profile in the horizontal channel at x=2.5~mm and y=2.5~mm against that of the alternative Navier Stokes setup (Fig.~\ref{fig:bnmrk_3d}b), which shows good agreement. Also overall, we can observe good agreement between the velocity profiles of the two approaches, when we compare Fig.~\ref{fig:bnmrk_3d_vel}a and  Fig.~\ref{fig:bnmrk_3d_vel}b. 
In Fig.~\ref{fig:bnmrk_3d_vel}c, we show the vectors of the fluid flow, in order to give a better overview of the merging of the two flow channels. We observe that, as the entire flow has to pass through a single channel, faster fluid flow velocities are present in the center of the fluid domain. We further note, that the FD approach can reproduce quite complex flow patterns, as seen in Fig.~\ref{fig:bnmrk_3d_vel}d, where we show the velocity vectors at the fracture intersection.


\subsection{Intersecting Fracture with rough surface topology}
\begin{figure*}
\begin{center}  
\includegraphics[width=0.95\textwidth]{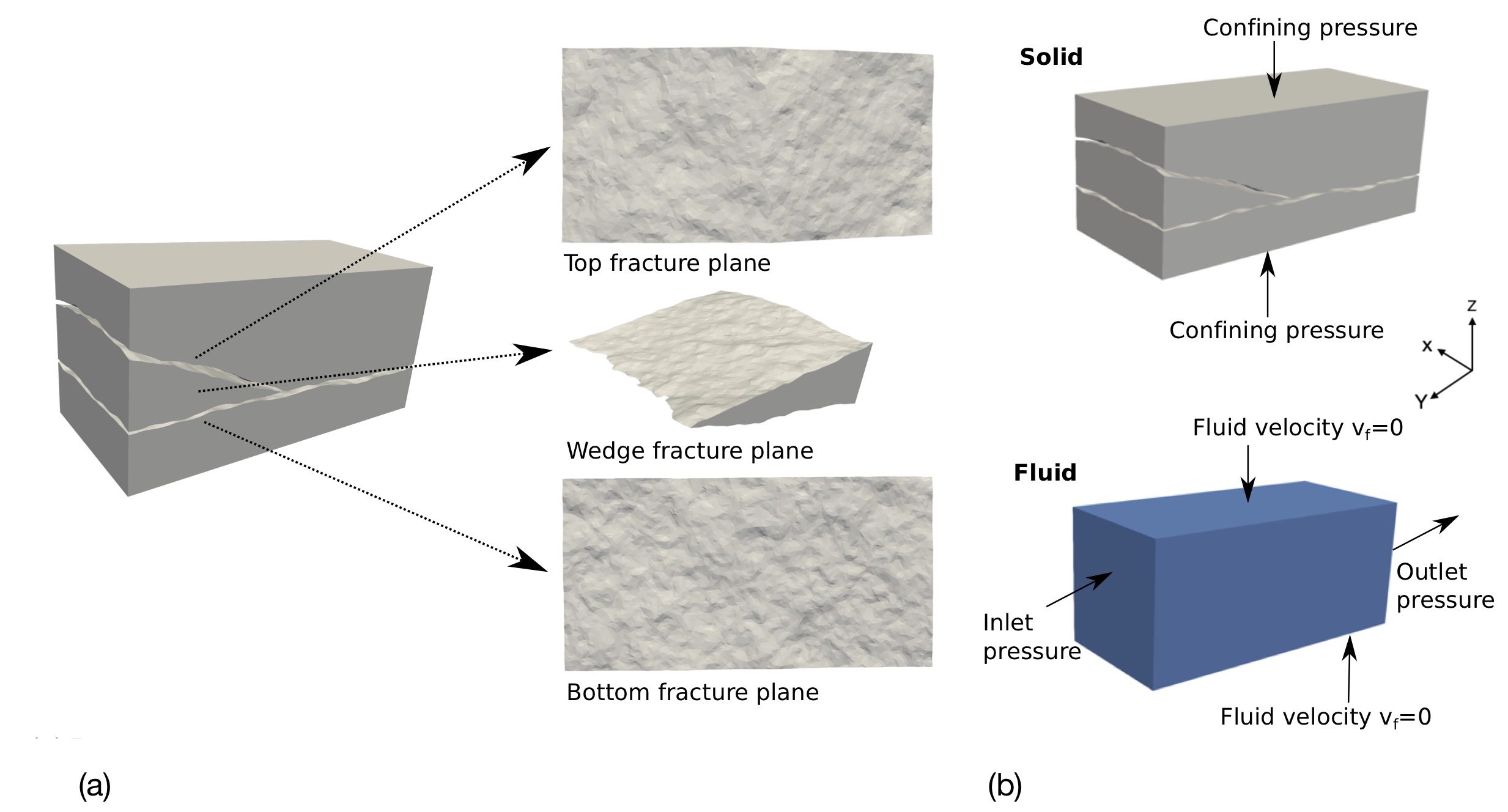}
\caption{System configuration: a) Fracture system configuration of the fluid and the three solid bodies (top fracture body, fracture wedge between top and bottom body, and bottom fracture body); and b) Boundary conditions for the solid (top) and the fluid (bottom).}
\label{fig:rf_geometries_bc}
\end{center}
\end{figure*}

The last numerical experiment demonstrates the capabilities of the FD method to simulate complex fracture intersection geometries, which are in this case three solid bodies with rough fracture surfaces (Fig.~\ref{fig:rf_geometries_bc}). Furthermore, in this set of experiments the fractures are in contact with each other, which results in more complex hydro-mechanical coupling scenarios. 
The contact is driven by increased loading of the solid bodies, leading to a decrease of the average fracture aperture.

\label{sec:rough_mesh}
The fracture intersection geometry in Fig.~\ref{fig:rf_geometries_bc}a consists of a main fracture and a smaller branch fracture. 
We generated the rough fracture surfaces with the software Synfrac \cite{Glover_1998a}.  Synfrac uses a spectral method to artificially generate surfaces that share the statistical properties, i.e. power spectral decomposition (PSD), observed in natural fractures. The surfaces are matching for longer wavelengths and nonmatching for shorter wavelengths. Roughness and mismatch are controlled by parameters.
The required input for fracture generation consists of the total fracture sizes of 100~mm$\times$100~mm, the fractal dimension of 2.5, an anisotropy ratio of 1.0 (resulting in isotropic fracture roughness), the standard deviation of the fracture roughness of 1, a transition length of 10~mm (i.e. the difference between macro and micro scales of the surface) and a mismatch length of 10~mm \cite{Glover_1998a}. 
The generated fractures showcase a smooth transition between matched and unmatched behaviour (fracture matching indicates the potential it has to match the opposite fracture wall when they are placed tightly next to each other), commonly encountered in actual rough-rock fractures.  
Since Synfrac outputs point clouds, these were transformed to a surface mesh in Meshlab \cite{meshlab_2008}, resampled to increase mesh density \cite{Igl17}, transformed into a solid body with the 3D CAD software Fusion~360 \cite{fusion} and finally imported in Trelis to generate a tetrahedral mesh of the solid body \cite{trelis}.

\begin{figure}[hbt!]
\begin{center}  
\includegraphics[width=0.95\textwidth]{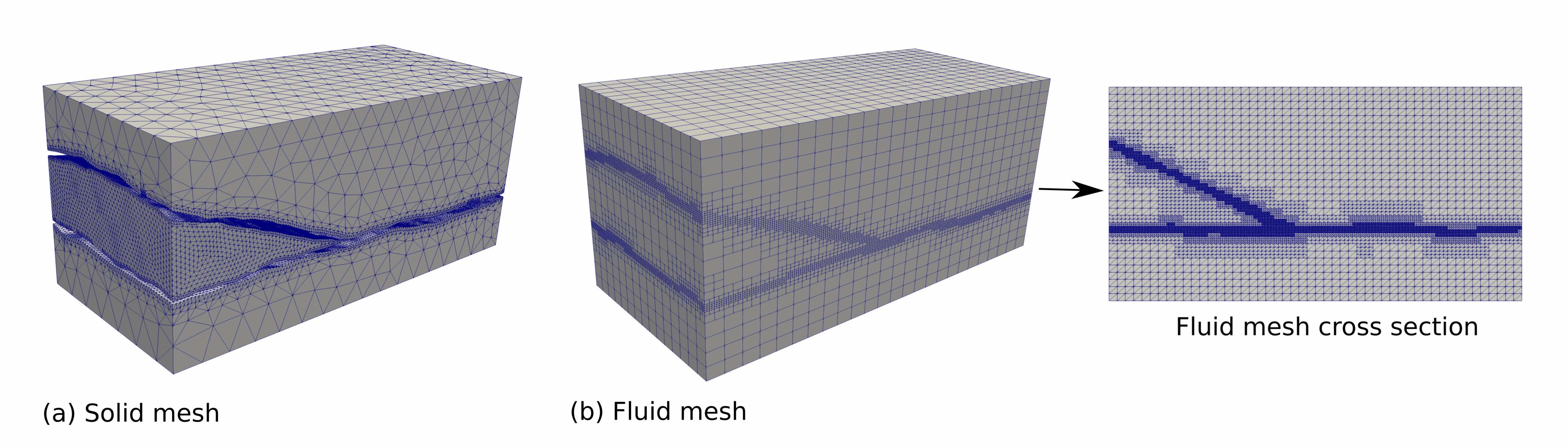}
\caption{Mesh geometries: a) Solid mesh and b) Refined fluid mesh.}
\label{fig:RF_Mesh}
\end{center}
\end{figure}

The size of the resulting solid mesh, consisting of three separate bodies, is 96$\times$44$\times$44~$\text{mm}^3$. The solid meshes of all bodies combined have 126'614 nodes and are all refined along the fracture surfaces to achieve higher resolutions at the fluid-solid boundaries (Fig.~\ref{fig:RF_Mesh}a). 
The fluid mesh has the same dimensions as the solid mesh, contains 1'433'019 nodes and is also refined in the fracture region (Fig.~\ref{fig:RF_Mesh}b). 

All  boundary conditions are shown in Fig.~\ref{fig:rf_geometries_bc}b. 
Notably, the system is compressed by applying displacements of 0.0 to 0.6~mm in the negative z-direction at the top and displacements of 0.0 to 0.6~mm in the z-direction at the bottom. As the fractures are initially in an open state, this results in increasing fracture closure over the course of the simulations.
For the fluid system, no-slip boundary conditions are applied at the top, bottom, front, and back boundaries, while fluid pressures of 1~MPa (on the left side) and 0.0~MPa (on the right side) yield fluid flow through the fractures from the left to the right.

\subsubsection{Fluid flow in fracture} \label{sec:roughFractureSurfaces}
\begin{figure*}[hbt!]
\begin{center}
\includegraphics[width=0.8\textwidth]{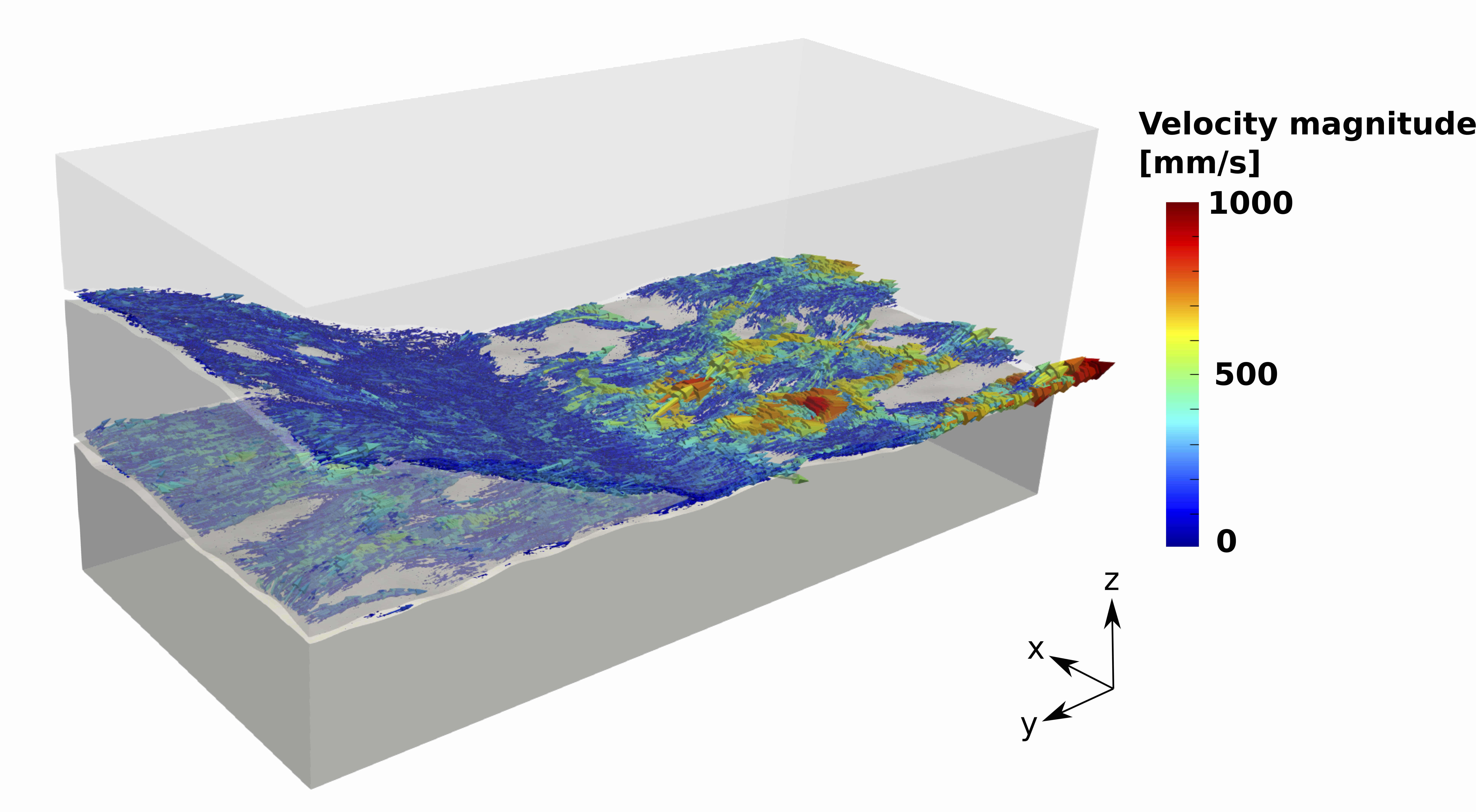}
\caption{Fluid field in the fractures under zero confining pressure.}
\label{fig:RF_vector}
\end{center}             
\end{figure*}

\begin{figure}[hbt!] 
\begin{center}
\includegraphics[width=0.45\columnwidth]{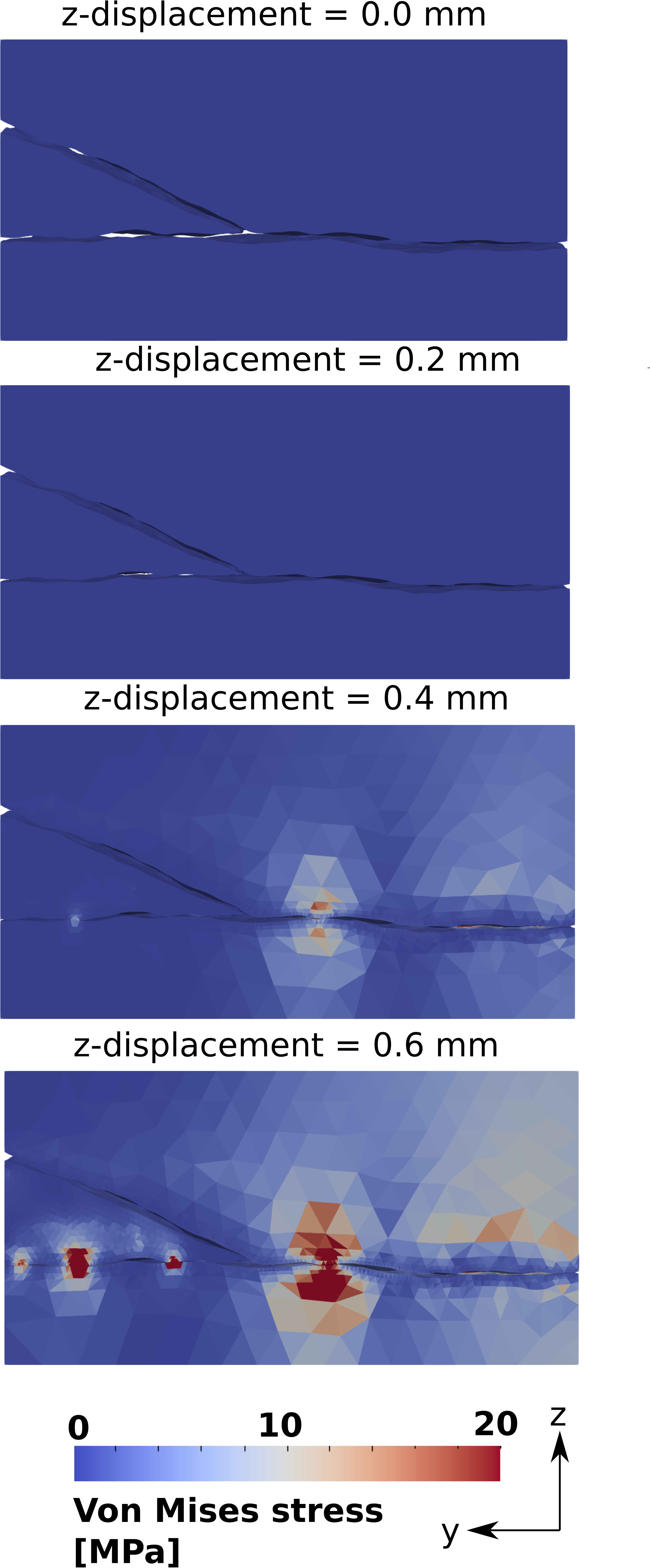}
\caption{
Von Mises stress distribution around the contact area under increasing confining pressure.}
\label{fig:RF_stress}
\end{center}
\end{figure}

Fig.~\ref{fig:RF_vector} depicts the fluid flow velocity vector field in the fractures for zero confining pressure. 
As can be expected from the previous example, the fluid flows through the two branches from the left side and merges into the long main fracture, where the fluid velocity magnitudes increase. Flow paths are more tortuous due to the rough fracture surfaces.
To observe the influences of increasing normal loads on the fracture geometry, we plot von Mises stresses inside the solid for different normal loading scenarios in Fig.~\ref{fig:RF_stress}. The magnitude of stresses focuses around the contact areas in the fractures and increases gradually during the loading process. This results in incrementally smaller apertures and increased contact areas, from which we expect that the area and magnitude of the fluid flow vector field also decreases.

\begin{figure*}[hbt!]
\begin{center}  
\includegraphics[width=0.95\textwidth]{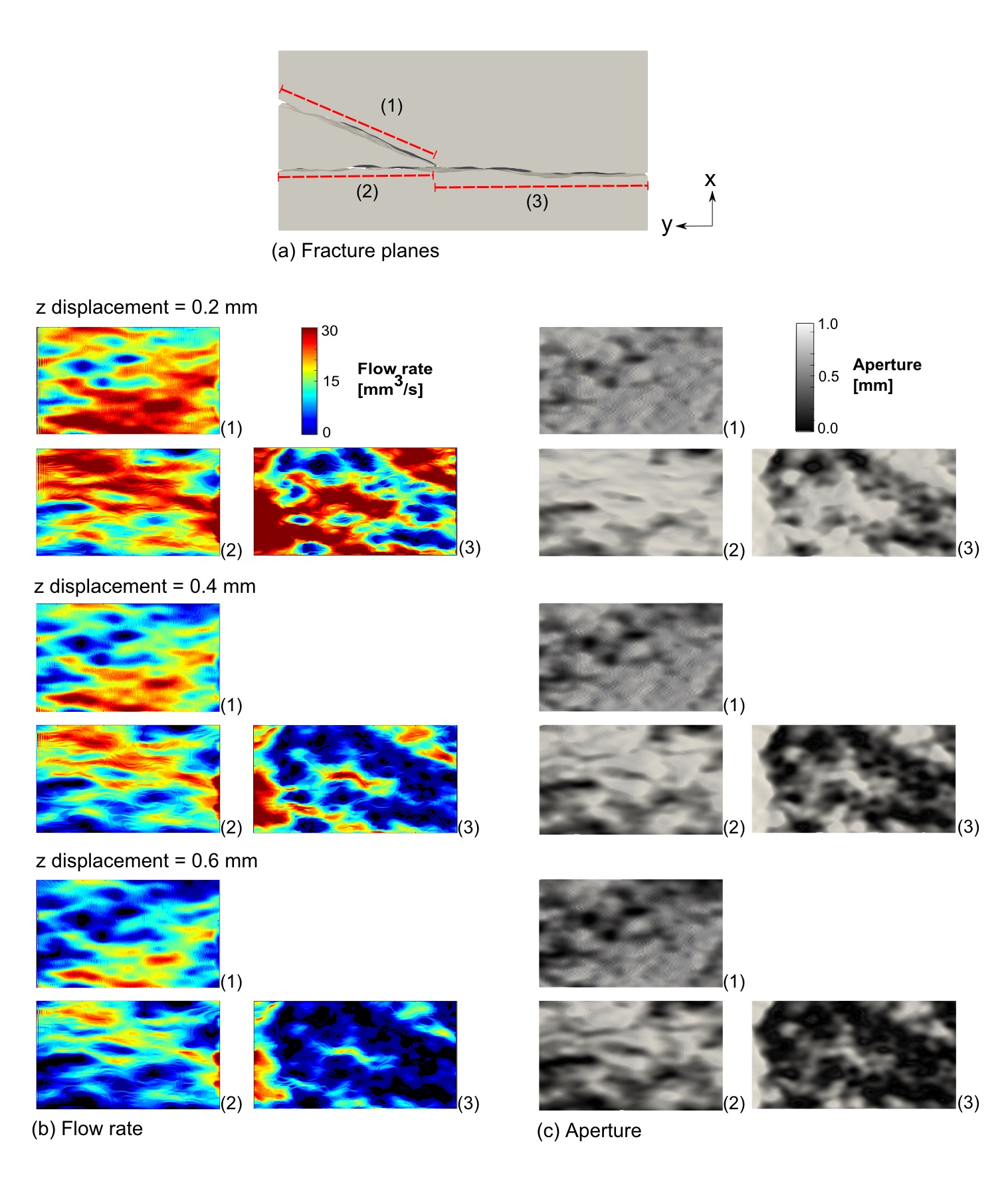}
\caption{
Fluid flow field and aperture distribution in the fracture intersection system, consisting of three fracture planes: 
a) Fracture plane labels with: (1) top short fracture, (2) bottom short fracture, and (3) long fracture; 
b) Volumetric fluid flow rates; and c) Aperture field.
}
\label{fig:RF_flowithgap}
\end{center}
\end{figure*}

To get a better understanding of the overall fluid flow in the system, we plotted aperture fields and flow rate magnitude in the three fractures in Fig.~\ref{fig:RF_flowithgap} at different stages of the loading process. 
To this end, we summed the fluid flow rates in the z-direction, which allows the visual observation of the flow field evolution with fracture closure (Fig.~\ref{fig:RF_flowithgap}b).
For comparison, we plot the respective aperture fields on the right of the flow fields in Fig.~\ref{fig:RF_flowithgap}c, where regions in contact, or almost in contact, are shown in black, while areas, where the fracture is open, are shown in white. 
As  expected, fluid flow concentrates in regions of large aperture, which become more scattered as the fractures close more. 
Merging of the fluid flow fields can be observed in the transition regions, where the right sides of fracture planes (1) and (2) merge with the left side of fracture plane (3). 
Since the whole system is subjected to symmetric closure from the top and bottom, the single fracture plane (Plane~3) closes notably more, compared to the two fracture planes, which are located above one another (Planes~1 and~2). 
This is caused by the similarity of the average apertures in all three fractures in the original system geometry. 
However, once the system is compressed by a finite solid displacement, this displacement is distributed to two fracture planes on the left side  and one fracture plane on the right side. 
This results in more closure in the single, long fracture on the right, while the shorter fracture planes only close partially.

\section{Conclusions}
In this article we have combined concepts of the fictitious domain method, modelling techniques for linear elastic contact, incompressible fluid flow, and variational transfer operators. We tested our fictitious domain method in a series of applications with increasing complexity to simulate fluid flow in intersecting fractures.
Variational transfer operators were employed for the transfer of physical variables between the solid and fluid, mapping the displacements between the non-matching surface meshes at the contact boundaries for the solid problem and for the generation of the fluid meshes. 

The results show, that the approach is capable of capturing fluid-solid boundaries of realistic fracture geometries with both complex fracture surface topographies, as well as complex fracture intersection configurations. 
This includes complex flow behaviour, such as merging of flow from two fractures into one fracture, flow channelling around contact regions or regions of low aperture and splitting of flow rates when fracture branches are encountered. 
The presented approach further demonstrated its ability to couple fluid flow and mechanical behaviour, which was tested by subjecting a complex fracture intersection geometry to increasing normal load. 
The latter constituted a multi-body contact problem, coupled with fluid flow. There, the increasing closure of the fracture planes and corresponding decrease of fracture aperture fields coincided with increased fluid flow channelling.

The presented application demonstrates that concepts of the fictitious domain method and variational transfer operators can be applied to geophysical problems.
The usage of transfer operators has allowed the coupling of different physical phenomena on different geometries, with the combination of solid and fluid mechanics. 
Thanks to the modularity of the presented approach, it can be extended to frictional contact, thermal processes, nonlinear materials or other physical processes, and thereby serves as a highly valuable tool for geophysical applications.

\section*{Acknowledgements}
We gratefully acknowledge funding by the Swiss Competence Center for Energy Research - Supply of Electricity (SCCER-SoE), by Innosuisse - Swiss Innovation Agency under Grant Number 28305.1 and the Swiss Federal Office of Energy (SFOE) under Grant Number SI/500676-02. The Werner Siemens-Stiftung (Foundation) is thanked for its support of the Geothermal Energy and Geofluids Group in the Department of Earth Sciences at ETH Zurich, Switzerland.

\bibliographystyle{plain} 
\bibliography{bibliography}
\end{document}